\documentclass[graybox]{article}

\usepackage{mathptmx}       % selects Times Roman as basic font
\usepackage{helvet}         % selects Helvetica as sans-serif font
\usepackage{courier}        % selects Courier as typewriter font
\usepackage{type1cm}        % activate if the above 3 fonts are
\usepackage{makeidx}         % allows index generation
\usepackage{graphicx}        % standard LaTeX graphics tool
\usepackage{multicol}        % used for the two-column index
\usepackage[bottom]{footmisc}% places footnotes at page bottom
\usepackage{epstopdf}
\makeindex             % used for the subject index
\begin{document}
	
	\title{Compartmental Spatial Multi-Patch Deterministic and Stochastic Models for Dengue}
	% Use \titlerunning{Short Title} for an abbreviated version of
	% your contribution title if the original one is too long
	\author{Wolfgang Bock and Yashika Jayathunga\\ TU Kaiserslautern\\ Technomathematics Group,\\ Department of Mathematics, \\P.O. Box 3049,\\ D-67653 Kaiserslautern ,\\ bock@mathematik.uni-kl.de,\\ yashika.jayathunga@gmail.com}
	%
	% Use the package "url.sty" to avoid
	% problems with special characters
	% used in your e-mail or web address
	%
	%\titlerunning{a}
	%\authorrunning{b}
	\maketitle

	\abstract{Dengue is a vector-borne viral disease increasing dramatically over the past years due to improvement in human mobility. The movement of host individuals between and within the patches are captured via a residence-time matrix. A system of ordinary differential equations (ODEs) modeling the spatial spread of disease among the multiple patches is used to create a system of stochastic differential equations (SDEs). Numerical
		solutions of the system of SDEs are compared with the deterministic solutions obtained via ODEs.}
	
	\section{Introduction}
	\label{sec:1}
	%Dengue fever which is a viral mosquito-borne infection  has become an international health problem in recent years as it leads to major cases of illness and death in the tropics and subtropics. About 390 million dengue cases causing 12,000 deaths per year are estimated according to WHO, see e.g.~\cite{WHO}. The virus is brought into to humans by infected female mosquitoes of the type Aedes aegypti. Infected humans carry and multiply the virus by transmitting it to uninfected mosquitoes.  Unprecedented population growth, unplanned urbanization, globalization ,  and latest faster modes of movement especially air transportation have been the major factors for the emergence and spreading of the diseases ~\cite{gubler2011dengue}. For several years, mathematical modeling is an important tool in epidemiology and dynamics. A series of different models for Dengue fever including stochastic and deterministic models, fractional differential equations, the effect of climate to the mosquito has been proposed, see e.g.~\cite{PGSW17,GABRSW16,  RMT14,GNSW13, AKMS12,AGS12CY01} and references therein.
	%
	%%Up to now, there is no effective treatment for dengue fever.  The main method to control and prevent the spread of the dengue virus is to fight vector mosquitoes through several measures including insecticides and preventing mosquitoes from accessing egg-laying habitats like in- and outdoor containers with stagnant water and reduce the bites by putting mosquito repellents.
	%\\ 
	With about 390 million dengue cases causing 12,000 deaths per year~\cite{WHO}, Dengue fever became an international health threat in many tropical and subtropical countries. 
	The disease dynamics are well known to be particularly complex with large fluctuations of disease incidences.

  The vectors are assumed to be located in the residing patch while the human population commutes between the different patches. Throughout the paper,  isolated areas of interest will be called as patches and in epidemiology, patches may refer to cities, countries, islands or health districts, like hospital districts and give a coarse spatial information about the disease spread.  In this paper, deterministic and stochastic models are presented and the 	mean numerical solutions of the stochastic model over large number of realizations are compared with the deterministic model dynamics. Throughout this work, only one strain of the Dengue virus is studied and further, studies can be extended to the rest of the virus serotypes ~\cite{normile2013surprising}. Note that the model can be without further adaption be applied to graphs.  
	
	The paper is organized as follows:	
	Following the introduction, deterministic and stochastic multi-patch models are described. Finally, two numerical examples are carried out to compare the deterministic and stochastic solutions of single patch model and two-patch model. 
	%%%% about the multistrain

	\section{ Deterministic Multi-patch Vector Host Models }
	\label{sec:2}
	In this section a  multi-patch dengue transmission Host-Vector model is introduced to examine the effect of host movement on dengue transmission dynamics. It is assumed that the patches will have a grid structure and the patch will be specified by the index $i$. There will be $n$ patches under consideration and the patches are coupled with the mobility of the hosts from one to the other.  An assumption is made that the vectors do not move between the patches  ~\cite{WHO, morlan1958urban, muir1998aedes} and the short term movement between the patches are coupled by a residence-budgeting time matrix $ P=(p_{ij})_{n \times n}$ for $i,j=1,2,...,n$, where $p_{ij} \in [0,1]$ and $\sum_{j=1}^{n}p_{ij}=1$ ~\cite{lee2015role}. The variables for the model corresponding to a specific patch denoted by $i$ are as follows: susceptible hosts \textbf{$S_i$}, infected hosts  \textbf{$I_i$}, and  infected vectors \textbf{$V_i$} which is an extension of the model  \cite{rochainfluence} and further, the model used here is an reduction of the model ~\cite{yashika1} assuming that the system is conservative. Both the host and vector populations ($N$ and $M$) are considered to be constant. Dengue fever is assumed to be transmitted by two means of interactions between host and vector: susceptible mosquitoes $(U_i)$ may interact with infected human $(I_i)$ individuals at a rate of $\frac{\vartheta_i}{N_i}$ and infected mosquitoes $(V_i)$ may interact with susceptible humans $(S_i)$ at a rate of $\frac{\beta_i}{M_i}$. The incidence rates at which humans and mosquitoes get infected are $\frac{\beta_i}{M_i} S_i V_i$ and $\frac{\vartheta_i}{N_i} U_i I_i$, respectively.  The birth and death rate of the humans is denoted by  $\mu_i$ ,  and it is assumed that the renovation of population happens in every 69 years ~\cite{unitednations2017}. The average time taken to recover is considered to be $\gamma_{i}$ and $\nu_i$ is the birth and death rate of the mosquitoes. Furthermore, it has been taken into account that the fact $S_i+I_i+R_i=N_i$ and $U_i+V_i=M_i$ and then the reduced  system of nonlinear ordinary differential equations for the $n$ number of patches are as follows:
	
	\begin{eqnarray}
	{\dot{\bf{S_i}}} &=& -S_i \sum_{j=1}^{n} \frac{\beta_j}{M_j} p_{ij}   V_j +\mu_{i} (N_i-S_{i})  \nonumber \\
	{\dot{\bf{I_i}}} &=&S_i \sum_{j=1}^{n} \frac{\beta_j}{M_j} p_{ij}   V_j-(\mu_{i}+\gamma_{i}) I_{i}\nonumber \\
	{\dot{\bf{V_i}}} &=&\frac{\vartheta_i}{N_i} (M_i-V_i)\sum_{j=1}^{n} p_{ji} I_j-\nu_{i} V_i 
	\label{SIRUV}	
	\end{eqnarray}
	
	The  basic reproductive number, $\mathbf{R}_{0}$ is the average number of secondary infectious cases when a single infectious individual is introduced into the whole susceptible population. The disease dies out if the basic reproduction number $\mathbf{R}_0<1$, and disease persists whenever $\mathbf{R}_0>1$~\cite{van2002reproduction}.The  local basic reproduction number for the patch specified by $i$ computed via the next-generation matrix  as described in~\cite{van2002reproduction} is $\mathbf{R}_{0i}=\sqrt{\frac{\beta_{i}\vartheta_{i}}{\nu_i(\gamma_{i}+\mu_{i})}}$.  For the simplicity, similar notation is used in the deterministic model and stochastic model. The formulation of   SDE models need defining three sets of random variables for $S_i,  I_i $ and $V_i$ where the dynamics of the corresponding variables depends on the relevant probabilities of the respective events. 
	
	\section{Stochastic Multi-patch Vector Host Models }
	\label{sec:3}
	Beyond an infectious disease outbreak, the population size and the state variables are massive resulting intensive computations in simulating the discrete-valued Markov chain process. In such a situation, the system of SDEs with continuous-valued random variables is used in order to approximate the discrete-valued process.  The random variables satisfy  $S_i,I_i,V_i \in [0,\infty]$. The system of stochastic differential equations is derived by the procedure stated in \cite{Allen}. The change of the system variables are denoted as $\Delta X(t)= (\Delta S_1(t),...,\Delta S_n(t),\Delta I_1(t),...,\Delta I_n(t),\Delta V_1(t),...,\Delta V_n(t))^T $  and the  system of ODEs given in \ref{SIRUV} are defined as $\frac{dX_k}{dt}=b_k$ for $k=1,2,...,3n$.

	Hence the SDE system takes the form $dX(t)=f(X(t))dt+B(X(t))dW(t)$ where B is a $3n \times  4n$ matrix which satisfies $B B^T=C$, where to order $\Delta t$,$C \Delta t$ is the approximate covariance matrix.  Further, $W(t)$ is a vector of $4n$ independent Wiener processes, corresponding to the $4n$ state transitions. Moreover, the system of Ito SDEs for the SIV model can be represented as,
	
	%\frac{}{}
	% \footnotesize{\begin{eqnarray}
	% 	B=\left( \begin{array}{rrrrrrrrrr}
	% 	-\sqrt{\frac{\beta SV}{M}} & \sqrt {\mu N} &-\sqrt{\mu S}&0&0&0&0&0&0&0 \\
	% 	\sqrt{\frac{\beta SV}{M}} &0&0& -\sqrt{\gamma I}&-\sqrt{\mu I} &0&0&0&0&0\\%2
	% 	0&0&0&\sqrt{\gamma I}&0&-\sqrt{\mu R}&0&0&0&0\\%3
	% 	0&0&0&0&0&0&\sqrt{\nu M }&-\sqrt{\nu U}&-\sqrt{\frac{\vartheta UI}{N}}&0\\%4
	% 	0&0&0&0&0&0&0&0&\sqrt{\frac{\vartheta UI}{N}}&-\sqrt{\nu V}
	% 	\end{array}\right) 
	% 	\end{eqnarray}}

	%\begin{eqnarray}
	%\mathrm{d}S&=&b_1\mathrm{d}t-\sqrt{\frac{\beta SV}{M}}\mathrm{d}W_1+\sqrt {\mu N} \mathrm{d}W_2 -\sqrt{\mu S}\mathrm{d}W_3\\ \nonumber %1
	%\mathrm{d}I&=&b_2\mathrm{d}t+\sqrt{\frac{\beta SV}{M}}\mathrm{d}W_1--\sqrt{\gamma I}\mathrm{d}W_4-\sqrt{\mu I}\mathrm{d}W_5\\\nonumber%2
	%\mathrm{d}R&=&b_3\mathrm{d}t+\sqrt{\gamma I}\mathrm{d}W_4-\sqrt{\mu R}\mathrm{d}W_6\\\nonumber %3
	%\mathrm{d}U&=&b_4\mathrm{d}t+\sqrt{\nu M }\mathrm{d}W_7-\sqrt{\nu U}\mathrm{d}W_8-\sqrt{\frac{\vartheta UI}{N}}\mathrm{d}W_9\\\nonumber
	%\mathrm{d}V&=&b_5\mathrm{d}t+\sqrt{\frac{\vartheta UI}{N}}\mathrm{d}W_9-\sqrt{\nu V}\mathrm{d}W_{10}\nonumber  \label{SDE model}
	%\end{eqnarray}
	%\begin{eqnarray}
	%{\dot{\bf{S_i}}} &=& -S_i \sum_{j=1}^{n} \frac{\beta_j}{M_j} p_{ij}   V_j +\mu_{i} (N_i-S_{i})  \nonumber \\
	%{\dot{\bf{I_i}}} &=&S_i \sum_{j=1}^{n} \frac{\beta_j}{M_j} p_{ij}   V_j-(\mu_{i}+\gamma_{i}) I_{i}\nonumber \\
	%{\dot{\bf{V_i}}} &=&\frac{\vartheta_i}{N_i} (M_i-V_i)\sum_{j=1}^{n} p_{ji} I_j-\nu_{i} V_i 
	%\label{SIRUV}	
	%\end{eqnarray}

	\begin{eqnarray}
	{\dot{\bf{S_i}}} &=&{ b_i -\sqrt{S_i \sum_{j=1}^{n} \frac{\beta_j}{M_j} p_{ij}   V_j } \mathrm{d}W_{6i-5}+\sqrt{\mu_{i} N_i}dW_{6i-4}-\sqrt{\mu_{i S_i}}\mathrm{d}W_{6i-3}} \nonumber \\
	{\dot{\bf{I_i}}} &=&{b_{i+n}+\sqrt{S_i \sum_{j=1}^{n} \frac{\beta_j}{M_j} p_{ij}   V_j} \mathrm{d}W_{6i-5}- \sqrt{(\mu_i+\gamma_i)I_i}dW_{6i-2}}\nonumber \\
	{\dot{\bf{V_i}}} &=&{b_{i+2n}+\sqrt{\frac{\vartheta_i}{N_i} (M_i-V_i)\sum_{j=1}^{n} p_{ji} I_j }\mathrm{d}W_{6i-1}-\sqrt{\nu_{i} V_i}\mathrm{d}W_{6i}}
	\label{modelwithcontrolreduced1}	
	\end{eqnarray}
	%\begin{align}
	%\mathrm{d}S&=b_1\mathrm{d}t-\sqrt{\beta_{vh} SV}\mathrm{d}W_1+\sqrt\mu_{hb}\mathrm{d}W_2 -\sqrt{\mu_{hd} S}\mathrm{d}W_3\\ \nonumber
	%\mathrm{d}I&=b_2\mathrm{d}t+\sqrt{\beta_{vh} SV}\mathrm{d}W_1-\sqrt{\gamma_{h}I}\mathrm{d}W_4-\sqrt{\mu_{hd}I}\mathrm{d}W_5\\\nonumber
	%\mathrm{d}R&=b_3\mathrm{d}t+\sqrt{\gamma_{h} I}\mathrm{d}W_4-\sqrt{\mu_{hd} R}\mathrm{d}W_6\\\nonumber
	%\mathrm{d}U&=b_4\mathrm{d}t+\sqrt{\mu_{vb} }\mathrm{d}W_7-\sqrt{\mu_{vd} U}\mathrm{d}W_8-\sqrt{\beta_{hv} UI}\mathrm{d}W_9\\\nonumber
	%\mathrm{d}V&=b_5\mathrm{d}t+\sqrt{\beta_{hv} UI}\mathrm{d}W_9-\sqrt{\mu_{vd} V}\mathrm{d}W_{10}\nonumber
	%\end{align}
	
	\section{Numerical Simulation}
	\label{sec:4}
	The dynamics of the deterministic model and SDE models are illustrated in two numerical examples one with single patch system where there is no host mobility and the other one with two patches, where the two patches are coupled via the residence time budgeting matrix. The numerical results for the single patch SDE model are derived via Euler-Maruyama scheme and Milstein scheme and the two patch model is numerically solved only by the Euler-Maruyama scheme up to now and comparing it with the results obtained via Milstein scheme is of future interest.
	
\begin{figure}[h]
%	\sidecaption
	% Use the relevant command for your figure-insertion program
	% to insert the figure file.
	% For example, with the graphicx style use
	\includegraphics[width=1\textwidth, height=0.8\textwidth]{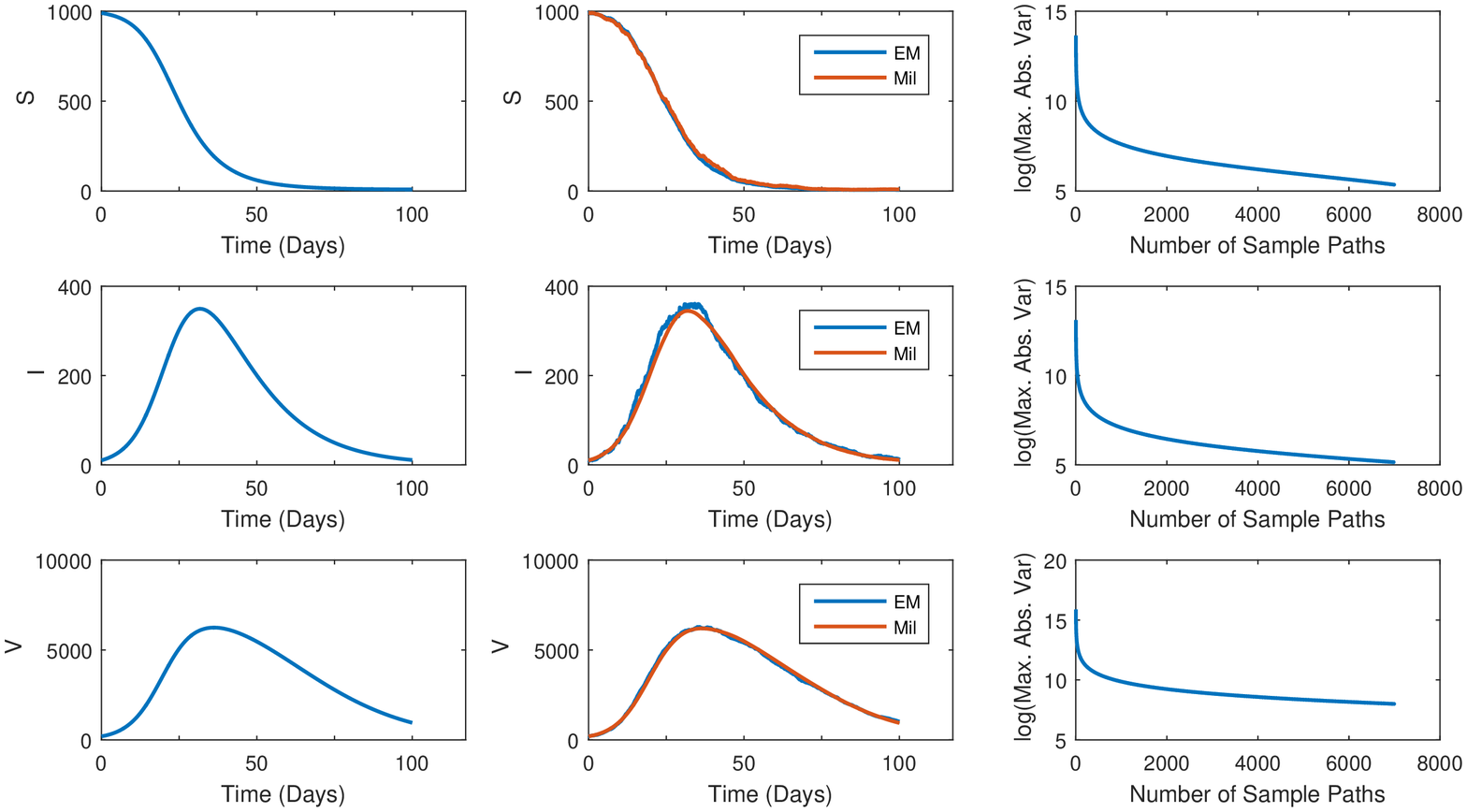}
	%
	% If no graphics program available, insert a blank space i.e. use
	%\picplace{5cm}{2cm} % Give the correct figure height and width in cm
	%
	\caption{Solution to the   model \ref{SIRUV} and \ref{modelwithcontrolreduced1} when $n=1$. The reproduction number of the single patch model is $\mathbf{R}_0=3.1614$. The initial susceptible hosts, infected hosts and infected vectors used in the simulations are 990, 10 and 200 respectively. The total number of  hosts and vectors are 1000 and 10000 respectively. Parameters used in the simulations are $\beta=1/7$,  $\mu=1/(69 \times 365)$, $\gamma=(1/14)$,  $\nu=1/10$, and $\vartheta=5\nu$.}
	\label{fig:1}       % Give a unique label
\end{figure}

\begin{figure}[h]
	%	\sidecaption
	% Use the relevant command for your figure-insertion program
	% to insert the figure file.
	% For example, with the graphicx style use
%	\includegraphics[width=1\textwidth]{figure2.eps}
	\includegraphics[width=1\textwidth, height=.8\textwidth]{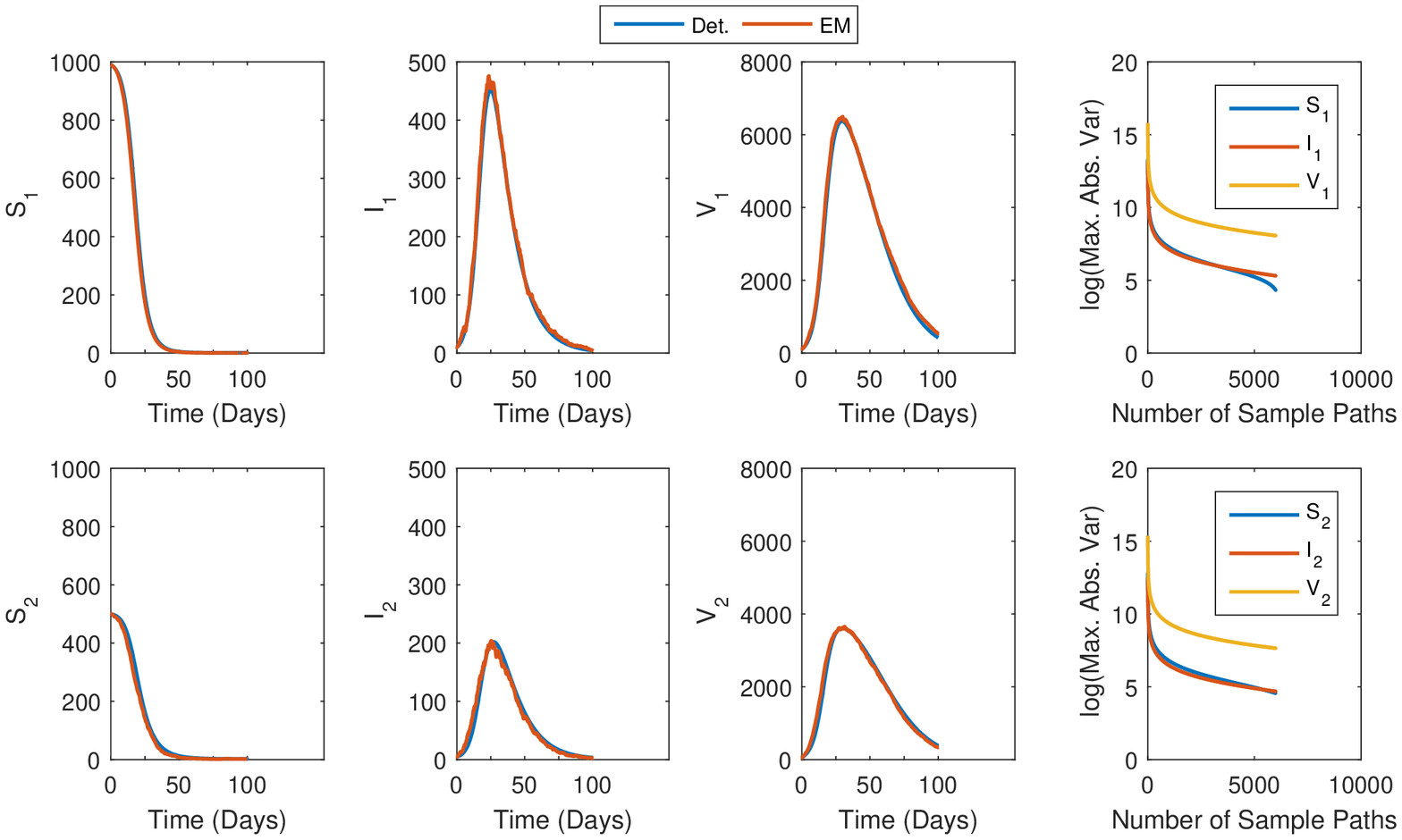}

	%
	% If no graphics program available, insert a blank space i.e. use
	%\picplace{5cm}{2cm} % Give the correct figure height and width in cm
	%
	\caption{Solution to the   model \ref{SIRUV} and \ref{modelwithcontrolreduced1} in the case of $ n=2$.  The initial susceptible hosts, infected hosts and infected vectors used in the simulations are given in the order of ($S_i, I_i, V_i=(990,500,10,5,100,50)$) respectively. The total number of  hosts and vectors are ($N_i, M_i=(1000,500)$). Parameters used in the simulations are $\beta_i=(2/7, 1/7)$,  $\mu_i=1/(69 \times 365)$, $\gamma_i=(1/14)$,  $\nu_i=1/10$, and $\vartheta_i=5\nu_i$.}
	\label{fig:2}       % Give a unique label
\end{figure}

	In Figure \ref{fig:1}, subplots shown by the first column refers to the deterministic solutions of the three state variables for $S$, $I$, and $V$ with reference to the single patch model where, subplots in the second column presents the mean of the numerical solutions of 7000 realizations  obtained by solving the SDE model  by using the Euler-Maruyama and Milstein schemes. Subplots shown by the third column  illustrates the logarithm of the maximum variance obtained via the  numerical scheme Euler-Maruyama,  against the number of sample paths.  It is clearly visible from the subplots in the third column that when the number of sample paths are higher the maximum variance of the solution gets gradually decreased.  
	
	Figure \ref{fig:2},  compares the mean solutions obtained via the Euler-Maruyama scheme to the stochastic model  (\ref{modelwithcontrolreduced1}) with the deterministic solutions for the two-patch model. Similar to the  single patch model here also, the  logarithm of the maximum variance of the average solutions reduce over the increment of number of sample paths.

	\section{Discussion}
	\label{sec:5}
	It can be concluded by the results obtained for the single patch model (Figure: \ref{fig:1}) that the solutions of the stochastic model is closely similar to the mean of  the deterministic solution with a higher number of sample paths and further, it has been confirmed by the variance plot where it illustrates that when the number of sample paths increase the variance of the solutions gets close enough to zero.  Also, the plot confirms that the solutions obtained via Milstein scheme gives more close results to the deterministic solution than the Euler-Maruyama scheme. 
	The movement of the hosts within the areas highly impact on the disease spread and here, it is investigated that the SDE model is closely related with the ODE model giving close results to the deterministic solutions.  In Figure: \ref{fig:2} the stochastic solutions are derived and compared with the deterministic solutions only via Euler-Maruyama scheme because there are some major difficulties in implementing the Milstein scheme to vector valued Wiener process, because the double stochastic integrals present in the scheme makes complexities with the implementation and will be of future interest to implement the two-patch model.  The results, which are present in this paper are for  $n=2$,  for better understanding but without any problem the simulations can be performed to a more generalized system with $n$ patches.

\end{document}